\documentclass[twocolumn,pra,superscriptaddress]{revtex4}

\usepackage{epsfig}
\usepackage{graphicx}

\newcommand{\Om}{{\Omega}}

\newcommand{\Ec}{{\cal E}}

\newcommand{\Dl}{{\Delta}}

\newcommand{\op}[2]{\left| #1 \right\rangle\left\langle #2 \right|}

\newcommand{\ket}[1]{\left| #1 \right\rangle}

\begin{document}


\title{
Nonlinear optics via double dark resonances
}

\author{S.~F.~Yelin}
\affiliation{Department of Physics, University of Connecticut, Storrs, CT 06269}
\author{V.~A.~Sautenkov}
\affiliation{Lebedev Physics Institute, Moscow, 117924, Russia}
\affiliation{Department of Physics, Texas A\&M University, College Station, TX 77843}
\author{M.~M.~Kash}
\affiliation{Department of Physics, Lake Forest College, Lake Forest, Illinois 60045}
\affiliation{Department of Physics, Texas A\&M University, College Station, TX 77843}
\author{G.~R.~Welch}
\affiliation{Department of Physics, Texas A\&M University, College Station, TX 77843}
\author{M.~D.~Lukin}
\affiliation{Department of Physics, Harvard University, Cambridge, MA 02138}

\date{\today}

\begin{abstract}
Double dark resonances originate from a coherent perturbation of a
system displaying electromagnetically induced transparency. We experimentally show and theoretically confirm that this leads to the
possibility of extremely sharp resonances prevailing even in the
presence of considerable Doppler broadening. A gas of
$^{87}$Rb atoms is subjected to a strong drive laser and a weak probe laser and a radio frequency field, where the
magnetic coupling between the Zeeman levels leads to nonlinear
generation of a comb of sidebands.
\end{abstract}


\maketitle

\section{Introduction}

Dark resonances arise from the quantum superposition states that are
decoupled from coherent and dissipative interactions. They are now a
well-known concept in quantum optics and laser spectroscopy, leading to
concepts such as electromagnetically induced transparency (EIT)
\cite{eit}, lasing without inversion and increased resonant
index of   
refraction \cite{lwi}, highly efficient resonant nonlinear optics
\cite{nlin,lukin00,hemmer95}, ``slow'' light \cite{hau99,kash99,budker99,koch00}, coherent storage of photon states \cite{liu01,phillips01,stop}, and femto-second generation \cite{harris98,hakuta92}.

As a rule, any perturbation of the dark state leads to undesirable
decoherence. However, it was shown theoretically that ``perturbing'' the dark state coherently
leads to multiple quantum superposition states that interact
coherently. These so-called ``double-dark'' states \cite{lukin99,ye02} can
be used to mitigate decoherence effects 
and thus enlarge the domain of dark state physics. The perturbation
can come from different sources: microwave or optical 
coherent fields, DC-fields, non-adiabatic optical coupling
(i.e. effective fields). In this paper we discuss theoretically and experimentally the use of the magnetic component of a radio
frequency (RF) field. We observe two sharp resonances at shifted frequencies. We also demonstrate that the double dark resonances allow to enhance the nonlinear optical process. In particular, we demonstrate the efficient generation of a comb of multiple frequency components associated with double dark resonances. 

\section{Theory}
\subsection{Dark resonances and electromagnetically induced transparency}

The principle of electromagnetically induced transparency (EIT) is the
destructive interference of two absorption channels, which originate
from a Stark split resonance. It
is based on a so-called {\it dark resonance}: 
\begin{figure}[ht]
\epsfxsize=2.5in
\centerline{\epsffile{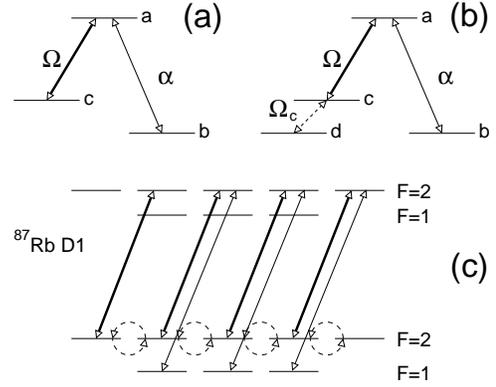}}
\caption{\protect\label{f_lambda}
(a) Energy level scheme of a $\Lambda$-system. (b) Extension of (a)
by an additional weak coherent field $\Om_c$ to a four-level system. (c) ``Four level'' system in $^{87}$Rb as used in the experiment.}
\end{figure}
For example, in a
$\Lambda$-system like the one in Fig.~\ref{f_lambda}a with the excited state
$\ket{a}$ coupled to the metastable states $\ket{b}$ and $\ket{c}$ via
strong driving and weak probe fields with Rabi frequencies $\Om$ and
$\alpha$, respectively, the 
resonant interaction Hamiltonian is $H=\hbar\Om\op{c}{a} +
\hbar\alpha\op{b}{a} + \text{H.c}.$. The superposition state
$\ket{+} 
\propto \Om\ket{c} + \alpha\ket{b}$ is coupled by both fields,
$H=\hbar\Om_+\op{+}{a}+\text{H.c.}$, with $\Om
_+^2=\Om^2+\alpha^2$, whereas the
orthogonal state $\ket{-}\propto \Om\ket{b}-\alpha\ket{c}$ is {\it
dark}, and $H\ket{-}=0$. This is true as long as the probe and driving
fields are in two-photon resonance with the transition
$\ket{b}\leftrightarrow\ket{c}$. If all the atoms in the medium are in the dark state
the medium is transparent to both fields
(see probe spectrum in Fig.~\ref{f_curves} (broken line)).
\begin{figure}[ht]
\epsfxsize=2in
\centerline{\epsffile{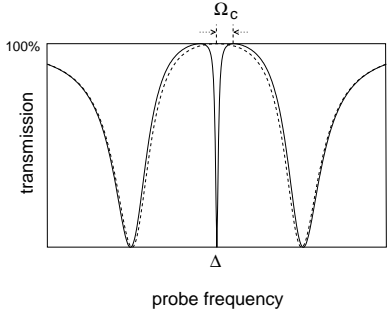}}
\caption{\protect\label{f_curves}
Transmission spectrum (dashed line) of probe field $\Ec$ of Fig.~\ref{f_lambda}a if the drive field $\Om$ is on resonance. The probe field is resonant at the center of the figure. The solid line depicts the same transmission spectrum in the presence of an additional weak coherent perturbation (cf.~Fig.~\ref{f_lambda}b). The transparency is split by $2\Om_c$.}
\end{figure}

\subsection{Double-dark resonances}

The usual mechanism by which  perturbing the dark state
destroys the transparency has been reviewed recently \cite{lukin99}: Here we show that a  coherent
perturbation, 
e.g., a weak monochromatic electromagnetic field with Rabi frequency
$\Om_c$ coupling an
additional state (see Fig.\ref{f_lambda}b), can display interesting and highly
coherent features. This case is depicted in Fig.~\ref{f_curves} (solid line), where
a sharp absorption line appears on resonance, that splits the point
of transparency in two. The two transmission maxima are  now detuned by $\pm\Om_c$. This absorption line  can be very narrow, as its width scales
with the ratio of the intensities $\Om_c^2/\Om^2$. Nevertheless, its
depth shows that the resonant absorption is as strong as that  of the original
Stark-split levels. For a simplified
explanation of this phenomenon, the dressed-state picture is very useful, where the Hamiltonian is diagonalized with respect to 
driving and perturbing fields. The
resulting resonant state, $\ket{0} \approx \ket{d} +
\Om_c/\Om\ket{a}$, has a small admixture of the excited state, and thus it
decays very slowly, which  results in a very narrow resonance. In this picture, it is easy to prove that in the all-resonant
case, detuning the field $\Om_c$ by $\Dl_{\rm RF}$ implies a frequency
shift of the novel resonance line by $\Dl_{\rm RF}$.

\subsection{Model}

\begin{figure}[h]
\epsfxsize=2.5in
\centerline{\epsffile{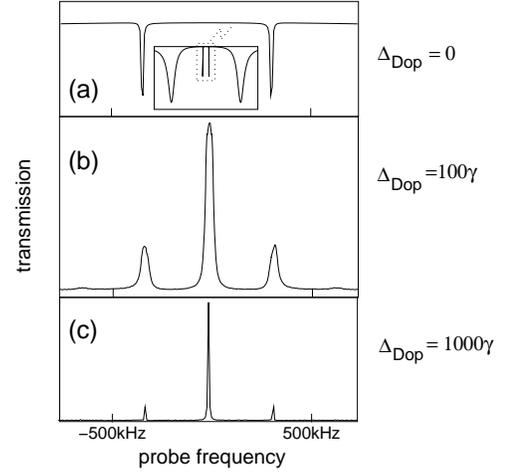}}
\caption{\protect\label{f_dopcurve}
(a) Transmission spectrum as in Fig.~\ref{f_lambda}b, but for a
situation in which $\ket{c}$ and $\ket{d}$ have the same energy, and the
additional field is a fixed-frequency RF field. The main figure is a
blowup of the inset. (b) Same as (a), but with a (realistic) Doppler
broadening of $\Dl_{\rm Dop}\sim 500$ MHz, which is approximately 100 natural linewidths
of the probe transition. (c) Same as in (b), but with ten times higher
Doppler broadening, in order to demonstrate the Doppler narrowing effect.}
\end{figure}

The experiment described in this article follows a slightly different
setup: The states $\ket{c}$ and $\ket{d}$ are degenerate (i.e., Zeeman
sublevels), whereas the
new field, a radio frequency (RF) field, has the fixed frequency
$\nu_{RF}$. This situation is the same as if there were two novel
fields that are blue and red detuned by $\pm\Dl_{\rm RF}=\nu_{\rm RF}$. One can thus
expect two novel lines with half the absorption depth, shifted by $\pm\nu_{RF}$ compared to the
center, as in Fig.~\ref{f_dopcurve}a (inset).
\begin{figure}[ht]
\epsfxsize=2.5in
\centerline{\epsffile{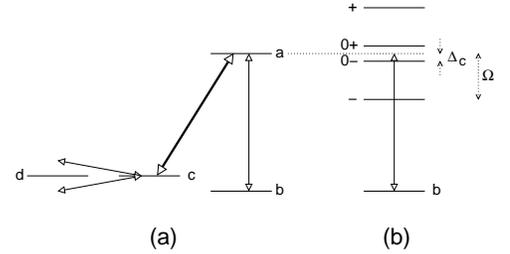}}
\caption{\protect\label{f_dressed} (a) Simple level system of the double dark setup as used in our experiment, where the RF field can be seen as red and blue detuned. (b) Same system, but in a dressed state picture. The level $\ket{d}$ is split into two by the red and blue detuned RF components, whereas the dressed states $\ket{+}$ and $\ket{-}$ are nearly the same as in a simple $\Lambda$-system.}
\end{figure}

In order to understand this behavior, we investigate briefly the dressed state system (i.e., the eigenvalues/eigenstates of the atom + drive + RF - system)(cf. Fig.~\ref{f_dressed}): 
The Hamiltonian of this setup (in the interaction picture without the probe field) is
\begin{eqnarray}
H &=& \hbar\Om\op{a}{c} + \frac{\hbar}{2}\Om_c\left( \op{d_1}{c}+\op{d_2}{c} \right) + H.c. \nonumber\\
&& + \hbar\Dl_{\rm RF}\left(\op{d_1}{d_1} -\op{d_2}{d_2} \right)\;.
\end{eqnarray}
Here, $\ket{d}$ is split into $\ket{d_1}$ and $\ket{d_2}$ to include the fact that the RF field is red and blue detuned at the same time. From this Hamiltonian we find the dressed state system 
\begin{eqnarray}
\ket{\pm} &\approx& \frac{1}{\sqrt{2}}\left(\ket{a}\pm\ket{c} \right)\\
&&\text{with}\nonumber\\
\lambda_\pm &\approx& \pm\Om\nonumber\\
&&\text{and}\nonumber\\
\ket{0\pm} &\approx& \ket{d_{1/2}} + \frac{\Om_c}{2\Om} f(\Om,\Dl_{\rm RF}) \left( \pm \ket{a} + \frac{\Dl_{\rm RF}}{\Om}\ket{c} \right)\\
&&\text{with}\nonumber\\
\lambda_{0\pm} &\approx& \pm \Dl_{\rm RF} \nonumber
\end{eqnarray}
Here we use $f(\Om,\Dl_{\rm RF}) = \Om^2/(\Om^2-\Dl_{\rm RF}^2)$ and assume $\Om\ne\Dl_{\rm RF}$. The small admixture of the excited state makes sure that the dressed states $\ket{0\pm}$, replacing $\ket{d}$, decay with a very narrow linewidth, and the eigenenergies coincide directly with the frequencies of the absorption lines.

\subsection{Nonlinear sideband generation}

\begin{figure}[ht]
\epsfxsize=2.5in
\centerline{\epsffile{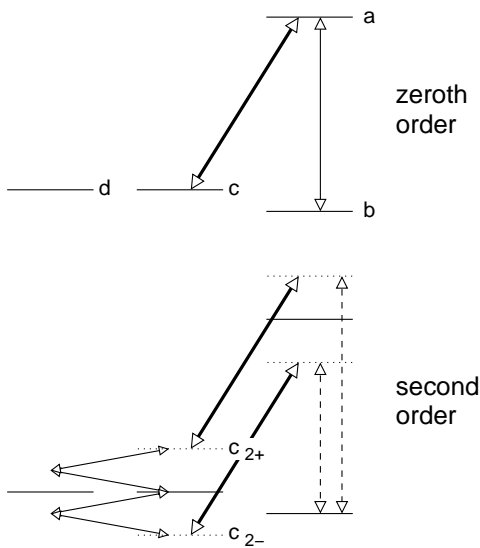}}
\caption{\protect\label{f_nonlinlevels}
Level structure illustrating how a non-resonant (i.e., finite-frequency) RF field can generate second order frequencies. The solid levels are pure atomic states, the dashed ones are virtual states created by the fields. The solid field arrows designate in- and outgoing fields, whereupon the dashed arrows signify fields that are created parametrically, and thus only outgoing.}
\end{figure}

This arrangement also produces efficient nonlinear modulation: sidebands appear in even multiples of the RF frequency $\nu_{\rm RF}$. 
This can be explained the following way:
Two RF photons couple (via state $\ket{d}$) to virtual states
$\ket{c_{2\pm}}$, that are lying exactly $2 \nu_{RF}$ above and below the
level of $\ket{c}$ (see Fig.~\ref{f_nonlinlevels}). The driving field parametrically creates a
transparency for the probe field at $2 \nu_{RF}$ above and below its
original frequency. Higher order nonlinearities then can be seen for
fourth, sixth, etc., order virtual levels $\ket{c_{4\pm}}$,
$\ket{c_{6\pm}}$, etc., with decreasing intensities (see section \ref{s_manynonlinears}).  

\subsection{Doppler narrowing}

\begin{figure}[h]
\epsfxsize=3.1in
\centerline{\epsffile{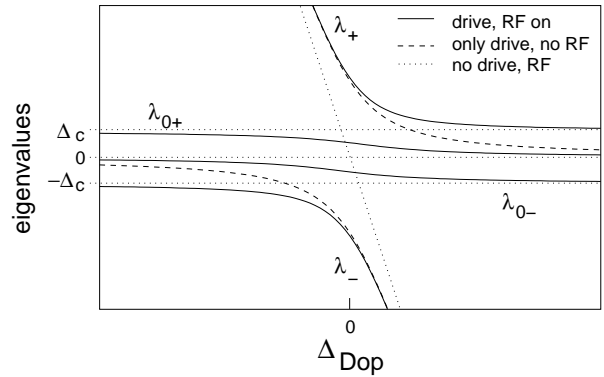}}
\caption{\protect\label{f_eigen}
Eigenvalues of the system as a function of the Doppler shift  $\Dl_{\rm Dop}$, i.e., for all different velocity groups.The eigenvalues in this figure depict the true energy level of the atomic states, and therefore would be at the center of any absorption line, whereas transmission would be visible somewhere between the eigenvalues. Dotted lines depict the energy levels of the atomic states when neither the drive ($\Om$) nor the RF field ($\Om_c$) is present. The broken lines are for $\Om_c=0$, but $\Om\ne0$. The solid lines give the solution for all three fields, RF, drive and probe, present.}
\end{figure}

The experiment is performed in a hot gas and therefore
Doppler broadening is an important consideration. We will give here a brief
explanation of the impact of Doppler shifts, as they pose an
interesting puzzle. 
For each velocity component of the atomic gas, the probe and drive transitions
are affected by Doppler shifts equally, since these shifts depend
linearly on the frequency and probe and drive fields copropagate. The
RF transition, however, is essentially unaffected because of
its very low frequency. It can be expected that the transparency on
resonance in the Doppler-broadened case gives rise to a transparency
whose width decreases for increasing
Doppler broadening. Surprisingly, there are Doppler
narrowed {\it transmission} lines replacing the absorption lines of
the Doppler free case. This ``Doppler narrowing'' can be explained
using Fig.~\ref{f_eigen}.

For each velocity group there are 4 eigenvalues in the
system. Each eigenvalue is responsible for an absorption line. In
the limit of vanishing RF field, and with the probe and drive fields on
resonance for the zero velocity group $\Dl_{\rm Dop}=0$, they are
\begin{eqnarray*}
\lambda_{0\pm}^{\rm Dop} &=& \pm\Dl_{\rm RF}\;,\\
\lambda_\pm^{\rm Dop} &=& \frac{\Dl_{\rm Dop}}{2} \pm \sqrt{ \frac{\Dl_{\rm
Dop}^2}{4} + \Om^2}\;,
\end{eqnarray*}
where $\Dl_{\rm Dop}$ is the Doppler frequency shift. Shown in Fig.~\ref{f_eigen} are the energy eigenvalues
$\lambda_\pm^{\rm Dop}$ and $\lambda_{0\pm}^{\rm Dop}$ of the atomic dressed states as functions of the 
Doppler shift $\Dl_{\rm Dop}$.
Eigenvalues $\lambda_\pm^{\rm Dop}$ (shown in the figure as broken lines) are responsible for the
Stark split absorption lines in usual EIT. The dotted lines show all eigenvalues without both, driving and RF fields ($\Om=\Om_c=0$. Note that in this case there are no avoided crossings.) With both fields  present 
all anticrossings appear in 
Fig.~\ref{f_eigen} (solid lines). Performing the Doppler average over
all components (which in mathematical terms means a Gaussian average over
all $\Dl_{\rm Dop}$)
leads to an absorption profile where narrow transparency lines can be
seen very close to the $\pm\Dl_{\rm RF}$ eigenvalues. The reason is obvious
from Fig.~\ref{f_eigen}, where the regions very close to $\pm\Dl_{\rm RF}$ and to 0 are the
only ones that are not crossed by any of the eigenvalues over the width
of all $\Dl_{\rm Dop}$. Thus these lines are nearly
at the frequency where there is maximum absorption in the Doppler free
case, and become narrower the broader the Doppler shift is spread. 

The resulting transmission curve can be seen in the simulation of
Fig.~\ref{f_dopcurve}, where the three lines, (one EIT and two double dark
lines) are clearly visible. The parameters in (b)
are chosen to coincide with the experimental reality: The Doppler
broadening in our experiment is 530 MHz.

The sideband generation is much more pronounced in the Doppler
broadened medium than in the non-broadened one, which seems surprising at first. It helps to remember
that the RF novel lines are absorption resonances in the
non-broadened case, whereas they are transmission resonances in the 
Doppler broadened case. These transmission lines coincide with the
zero-order dark resonance lines of the sideband spectra (see, e.g., Fig.~\ref{f_stepcurves}). The consequence is that the
ratio of neighboring sideband intensities $I_{4\Dl_{\rm RF}}/I_{2\Dl_{\rm RF}}
\approx I_{2\Dl_{\rm RF}}/I_0$ is roughly two orders of magnitude
higher for Doppler broadened media, such that the three-peak structure of
the fundamental mode can also be seen in the second order
sidebands (Fig.~\ref{f_stepcurves}(a,b)).

\section{Experiment}
\subsection{Setup}

\begin{figure}[h]
\epsfxsize=3.5in
\centerline{\epsffile{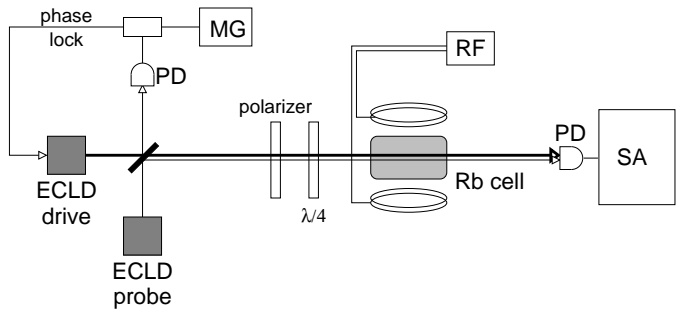}}
\caption{\protect\label{f_exp} Experimental setup. The drive and probe fields are produced by phase-locked extended cavity laser diodes (ECLD) and $\sigma_+$-polarized by a polarizer and $\lambda/4$-wave plate. A spectrum analyzer (SA) examines the probe--drive beat note. The weak coupling field is produced by a radio frequency (RF) coil.}
\end{figure}

The experiment was performed on the D$_1$ resonance line of $^{87}$Rb
(nuclear spin I=3/2, wavelength 795nm). The experimental setup is
shown in Fig.~\ref{f_exp}. In our experiment, external cavity diode lasers (ECDL)
were used for both the drive and the probe fields. The two collinear beams, both
of which are circularly polarized and which have the same
diameter of 0.1 cm are overlapped and directed into a glass cell of
length $l=4$ cm, filled with rubidium vapor. The vapor
consisted of $4\times 10^{11}$cm$^{-3}$ ($T=65^o$C) of the natural abundance of
$^{85}$Rb and $^{87}$Rb, plus 3 Torr of Ne buffer gas. The buffer gas is
needed in order to reduce the relaxation rate of the ground state
coherence, which for vapor densities of this order stems mostly from
the Rb atoms entering into and leaving the interaction region with the
laser beams. Thus the ground state decoherence rate depends on the
diameter of the laser beams and in our case can be estimated to be of
the order of 5 kHz. The cell was mounted inside a three-layered
magnetic shield. Perpendicular to the laser beams the RF field could be applied, using transverse
coils. The amplitude of the magnetic component of the RF field is chosen such that it is of the order 0.1 G. The
cell was heated using DC current in a high-resistance wire wound on the
outside of the middle shield. The atomic density can then be estimated
from the temperature in the cell using well-known vapor-pressure
curves, e.g. from \cite{nesm63}. The drive laser was tuned to the $5S_{1/2}
(F=2) \leftrightarrow 5P_{1/2} (F=2)$ transition; the probe laser was tuned to the $5S_{1/2}
(F=1) \leftrightarrow 5P_{1/2} (F=2)$ transition. This configuration of drive
and probe fields creates an appropriate $\Lambda$-system within the
$^{87}$Rb D1 manifold (see Fig.~\ref{f_lambda}c).  

There must be  minimal relative laser jitter
between probe and drive fields, so the
probe laser was phase-locked to the driving field frequency with an
offset frequency of 6.8 GHz, which is equal to the hyperfine ground state splitting between the $5S_{1/2}
(F=1)$ and $5S_{1/2} (F=2)$ states. This offset is fixed by a tunable
microwave frequency synthesizer.  
After passing through the cell the laser beams hit a fast
photodiode. The beat signal from this photodiode is
monitored by a microwave spectrum analyzer, which is used as a
broadband filter-amplifier with a bandwidth of 3 MHz, centered at the
hyperfine splitting frequency.

\subsection{Spectrum}

The lineshape of the dark resonances are then determined by scanning
the probe laser frequency, which was accomplished by tuning the
frequency of the synthesizer. While the laser power of the driving
field can be changed from 0 to 6 mW, the probe laser was held at 1\% of
the drive power. With these parameters a broadening of the dark
resonances could be observed with a slope of 30 kHz/mW. Field broadening
and density narrowing of the dark resonance were investigated in
Ref.~\cite{saut99}. In order to observe EIT resonance and
pronounced double dark structure, the driving field had to be strong enough to make the the additional
double dark structures lie well within the EIT resonance. (As was tested
experimentally in this scope, the novel features are not visible
otherwise.) The optimal working point thus was determined to be 2.2 mW
for the drive laser.
\begin{figure}[h]
\epsfxsize=3in
\centerline{\epsffile{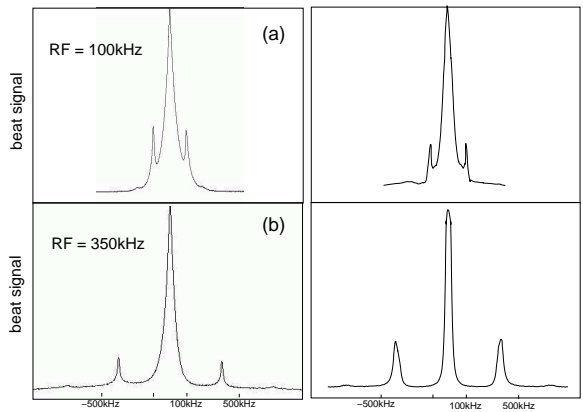}}
\caption{\protect\label{f_simplecurves} Transmission curves for two different RF frequencies, experiment
(right) and theory (left). In (a) the RF is 100 kHz, where the new
resonances still overlap with the dark resonance peak, whereas in (b),
with a RF of 350 kHz, these lines are well resolved. The RF magnetic field component 
in these curves was 20 mG in (a) and 80 mG in (b). }
\end{figure}

The double
dark resonance structures can be seen in
Fig.~\ref{f_simplecurves}a. These additional resonances are induced by
the alternating magnetic field with a frequency of 100 kHz. An RF field with a frequency of 350 kHz is shown in Fig.~\ref{f_simplecurves}b. The new transparency lines are well outside the EIT
peak. As can be seen in the figures, the
frequency interval between the dark resonance and the new resonances indeed
equals the RF frequency. The width of the novel resonances is 20 kHz,
which suggests a total decoherence rate between the three ground
states $\ket{b}$, $\ket{c}$, and $\ket{d}$ of 20 kHz.  

\subsection{Sideband generation}
\label{s_manynonlinears}

\begin{figure}[h]
\epsfxsize=3in
\centerline{\epsffile{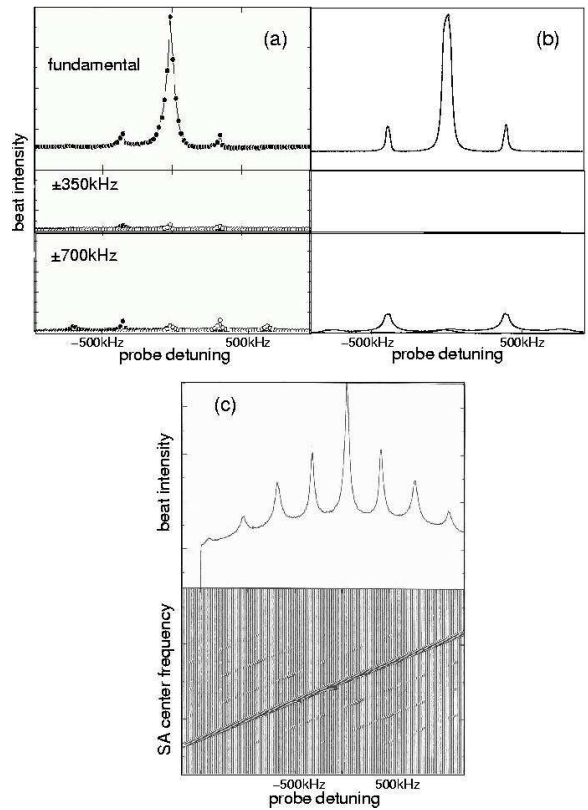}}
\caption{\protect\label{f_stepcurves} 
(a) Spectrum of the drive-probe beat note transmission. In the upper picture, the spectrum analyzer (SA) is centered around the fundamental, i.e., resonance frequency. In the two lower graphs, the SA is tuned to the first and second order sidebands, respectively. It is clear from these curves that only even
order sidebands are supported. (b) Simulations of curves in (a). Note that while in theory there are no first order sidebands, thee are small ones in the experimental data. (The reason for that is a broad fundamental mode.) (c) The upper curve is similar to the one in (a), but with much stronger RF fields (160 mG) such that multiple sidebands are visible. This spectrum is an integration over all SA frequencies. The resolution by SA frequencies is given in the lower picture.}
\end{figure}

The spectral distribution of the transmitted laser light was also
studied. For this purpose the spectral analyzer was switched to a
narrow band regime (bandwidth 30 kHz) and the central frequency of the
bandwidth was scanned. The laser frequency during the spectrum
analyzer scan was fixed.  In this procedure several new optical
components (sidebands) can be observed, which are separated by
frequency intervals equal to harmonics of the applied magnetic field
frequency. Obviously these optical components are generated during
the propagation of laser light through the coherently prepared Rb vapor. In order to
record their dependence on the laser
frequency, we present snapshots where frequency steps of 20 kHz are
taken, see Fig.~\ref{f_stepcurves}.

The fundamental frequency here is the
frequency of the beat-note of the drive- and probe- laser beams
without the cell. The signal at the fundamental frequency is similar
to the broadband detected signal in Fig.~\ref{f_simplecurves}b, as could
be expected. One can see the different behavior of the odd and even
harmonics (sidebands). The amplitude of the second order sidebands that are
observed at $\pm$700 kHz is higher than the the amplitude of the
first order sidebands that are observed at $\pm$ 350 kHz. The reason that
they are seen at all stems from the fact that the
fundamental mode is broad enough to seep into the first order
sidebands. In this 
case, the virtual states
$\ket{d_\pm}$, detuned by $\pm\Dl_{\rm RF}$ from state $\ket{d}$, are
directly coupled to $\ket{a}$, which leads to first order sidebands.
Note that in the simulations (Fig.~\ref{f_dopcurve}) no first order sidebands
are present. 

The three-peak structure in the main diagonal of the lower curves of
Fig.~\ref{f_stepcurves}(c) is the main contribution to the integrated
spectrum (upper curve), and for low RF intensities is all that is
seen (e.g., in (a,b)). In reality, however, this contribution makes up
only the fundamental mode. Sidebands originate in the
resonances with $\ket{c_{2\pm}}$, $\ket{c_{4\pm}}$, etc., and can be
seen for stronger RF intensity, such as in
Fig.~\ref{f_stepcurves}(c). There, each of the diagonals, which are offset from their neighbors
by $\pm2\nu_{\rm RF}$, consists of the fundamental three-peak
structure. Finally, in the integrated picture they add up to the
multi-peak spectrum in the upper curve.

\section{Conclusion}

In conclusion, we have experimentally demonstrated novel coherent
resonances as predicted in Ref.~\cite{lukin99}. These novel resonances
are very sharp and highly sensitive to frequency and intensity of a RF
field, which perturb the dark resonance. In addition, because of higher-order coupling, efficient non-linear sideband generation was
demonstrated. This scheme can be used in applications like sensitive nonlinear measurements such as Ref.~\cite{yelin02} and UV laser generation \cite{fry00}.

We want to thank M.~O.~Scully for his interest in this project and discussions leading to it, and we want to thank the Office of Naval Research and the National Science Foundation for support. SFY also acknowledges a Feodor-Lynen Fellowship from the Humboldt Foundation.

\bibliography{vol_xxx}

\begin{thebibliography}{20}
\expandafter\ifx\csname natexlab\endcsname\relax\def\natexlab#1{#1}\fi
\expandafter\ifx\csname bibnamefont\endcsname\relax
  \def\bibnamefont#1{#1}\fi
\expandafter\ifx\csname bibfnamefont\endcsname\relax
  \def\bibfnamefont#1{#1}\fi
\expandafter\ifx\csname citenamefont\endcsname\relax
  \def\citenamefont#1{#1}\fi
\expandafter\ifx\csname url\endcsname\relax
  \def\url#1{\texttt{#1}}\fi
\expandafter\ifx\csname urlprefix\endcsname\relax\def\urlprefix{URL }\fi
\providecommand{\bibinfo}[2]{#2}
\providecommand{\eprint}[2][]{\url{#2}}

\bibitem[{\citenamefont{Harris}(1997)}]{eit}
\bibinfo{author}{\bibfnamefont{S.}~\bibnamefont{Harris}},
  \bibinfo{journal}{Physics Today} \textbf{\bibinfo{volume}{50}},
  \bibinfo{pages}{36} (\bibinfo{year}{1997}), \bibinfo{note}{for review}.

\bibitem[{\citenamefont{Scully and Zubairy}(1997)}]{lwi}
\bibinfo{author}{\bibfnamefont{M.~O.} \bibnamefont{Scully}} \bibnamefont{and}
  \bibinfo{author}{\bibfnamefont{M.~S.} \bibnamefont{Zubairy}},
  \emph{\bibinfo{title}{Quantum optics}} (\bibinfo{publisher}{Cambridge,
  University Press, Cambridge}, \bibinfo{year}{1997}), \bibinfo{note}{for
  references}.

\bibitem[{\citenamefont{Merriam et~al.}(2000)\citenamefont{Merriam, Sharpe,
  Shverdin, Manuszak, Yin, and Harris}}]{nlin}
\bibinfo{author}{\bibfnamefont{A.~J.} \bibnamefont{Merriam}},
  \bibinfo{author}{\bibfnamefont{S.~J.} \bibnamefont{Sharpe}},
  \bibinfo{author}{\bibfnamefont{M.}~\bibnamefont{Shverdin}},
  \bibinfo{author}{\bibfnamefont{D.}~\bibnamefont{Manuszak}},
  \bibinfo{author}{\bibfnamefont{G.~Y.} \bibnamefont{Yin}}, \bibnamefont{and}
  \bibinfo{author}{\bibfnamefont{S.~E.} \bibnamefont{Harris}},
  \bibinfo{journal}{Phys.~Rev.~Lett.} \textbf{\bibinfo{volume}{84}},
  \bibinfo{pages}{5308} (\bibinfo{year}{2000}).

\bibitem[{\citenamefont{Lukin and Imamo\u{g}lu}(2000)}]{lukin00}
\bibinfo{author}{\bibfnamefont{M.~D.} \bibnamefont{Lukin}} \bibnamefont{and}
  \bibinfo{author}{\bibfnamefont{A.}~\bibnamefont{Imamo\u{g}lu}},
  \bibinfo{journal}{Phys.~Rev.~Lett.} \textbf{\bibinfo{volume}{84}},
  \bibinfo{pages}{1419} (\bibinfo{year}{2000}).

\bibitem[{\citenamefont{Hemmer et~al.}(1995)\citenamefont{Hemmer, Katz,
  Donoghue, Cronin-Golomb, Shahriar, and Kumar}}]{hemmer95}
\bibinfo{author}{\bibfnamefont{P.~R.} \bibnamefont{Hemmer}},
  \bibinfo{author}{\bibfnamefont{D.~P.} \bibnamefont{Katz}},
  \bibinfo{author}{\bibfnamefont{J.}~\bibnamefont{Donoghue}},
  \bibinfo{author}{\bibfnamefont{M.}~\bibnamefont{Cronin-Golomb}},
  \bibinfo{author}{\bibfnamefont{M.~S.} \bibnamefont{Shahriar}},
  \bibnamefont{and} \bibinfo{author}{\bibfnamefont{P.}~\bibnamefont{Kumar}},
  \bibinfo{journal}{Opt.~Lett.} \textbf{\bibinfo{volume}{20}},
  \bibinfo{pages}{982} (\bibinfo{year}{1995}).

\bibitem[{\citenamefont{Hau et~al.}(1999)\citenamefont{Hau, Harris, Dutton, and
  Behroozi}}]{hau99}
\bibinfo{author}{\bibfnamefont{L.~V.} \bibnamefont{Hau}},
  \bibinfo{author}{\bibfnamefont{S.~E.} \bibnamefont{Harris}},
  \bibinfo{author}{\bibfnamefont{Z.}~\bibnamefont{Dutton}}, \bibnamefont{and}
  \bibinfo{author}{\bibfnamefont{C.~H.} \bibnamefont{Behroozi}},
  \bibinfo{journal}{Nature} \textbf{\bibinfo{volume}{397}},
  \bibinfo{pages}{594} (\bibinfo{year}{1999}).

\bibitem[{\citenamefont{Kash et~al.}(1999)\citenamefont{Kash, Sautenkov,
  Zibrov, Hollberg, Welch, Lukin, Rostovtsev, Fry, and Scully}}]{kash99}
\bibinfo{author}{\bibfnamefont{M.~M.} \bibnamefont{Kash}},
  \bibinfo{author}{\bibfnamefont{V.~A.} \bibnamefont{Sautenkov}},
  \bibinfo{author}{\bibfnamefont{A.~S.} \bibnamefont{Zibrov}},
  \bibinfo{author}{\bibfnamefont{L.}~\bibnamefont{Hollberg}},
  \bibinfo{author}{\bibfnamefont{G.~R.} \bibnamefont{Welch}},
  \bibinfo{author}{\bibfnamefont{M.~D.} \bibnamefont{Lukin}},
  \bibinfo{author}{\bibfnamefont{Y.}~\bibnamefont{Rostovtsev}},
  \bibinfo{author}{\bibfnamefont{E.~S.} \bibnamefont{Fry}}, \bibnamefont{and}
  \bibinfo{author}{\bibfnamefont{M.~O.} \bibnamefont{Scully}},
  \bibinfo{journal}{Phys.~Rev.~Lett.} \textbf{\bibinfo{volume}{82}},
  \bibinfo{pages}{5229} (\bibinfo{year}{1999}).

\bibitem[{\citenamefont{Budker et~al.}(1999)\citenamefont{Budker, Kimball,
  Rochester, and Yashchuk}}]{budker99}
\bibinfo{author}{\bibfnamefont{D.}~\bibnamefont{Budker}},
  \bibinfo{author}{\bibfnamefont{D.~F.} \bibnamefont{Kimball}},
  \bibinfo{author}{\bibfnamefont{S.~M.} \bibnamefont{Rochester}},
  \bibnamefont{and} \bibinfo{author}{\bibfnamefont{V.~V.}
  \bibnamefont{Yashchuk}}, \bibinfo{journal}{Phys.~Rev.~Lett.}
  \textbf{\bibinfo{volume}{83}}, \bibinfo{pages}{1767} (\bibinfo{year}{1999}).

\bibitem[{\citenamefont{Kocharovskaya et~al.}(2001)\citenamefont{Kocharovskaya,
  Rostovtsev, and Scully}}]{koch00}
\bibinfo{author}{\bibfnamefont{O.}~\bibnamefont{Kocharovskaya}},
  \bibinfo{author}{\bibfnamefont{Y.}~\bibnamefont{Rostovtsev}},
  \bibnamefont{and} \bibinfo{author}{\bibfnamefont{M.~O.}
  \bibnamefont{Scully}}, \bibinfo{journal}{Phys.~Rev.~Lett.}
  \textbf{\bibinfo{volume}{86}}, \bibinfo{pages}{628} (\bibinfo{year}{2001}).

\bibitem[{\citenamefont{Liu et~al.}(2001)\citenamefont{Liu, Dutton, Behroozi,
  and Hau}}]{liu01}
\bibinfo{author}{\bibfnamefont{C.}~\bibnamefont{Liu}},
  \bibinfo{author}{\bibfnamefont{Z.}~\bibnamefont{Dutton}},
  \bibinfo{author}{\bibfnamefont{C.~H.} \bibnamefont{Behroozi}},
  \bibnamefont{and} \bibinfo{author}{\bibfnamefont{L.~V.} \bibnamefont{Hau}},
  \bibinfo{journal}{Nature} \textbf{\bibinfo{volume}{409}},
  \bibinfo{pages}{490} (\bibinfo{year}{2001}).

\bibitem[{\citenamefont{Phillips et~al.}(2001)\citenamefont{Phillips,
  Fleischhauer, Mair, Walsworth, and Lukin}}]{phillips01}
\bibinfo{author}{\bibfnamefont{D.}~\bibnamefont{Phillips}},
  \bibinfo{author}{\bibfnamefont{A.}~\bibnamefont{Fleischhauer}},
  \bibinfo{author}{\bibfnamefont{A.}~\bibnamefont{Mair}},
  \bibinfo{author}{\bibfnamefont{R.}~\bibnamefont{Walsworth}},
  \bibnamefont{and} \bibinfo{author}{\bibfnamefont{M.~D.} \bibnamefont{Lukin}},
  \bibinfo{journal}{Phys.~Rev.~Lett.} \textbf{\bibinfo{volume}{86}},
  \bibinfo{pages}{783} (\bibinfo{year}{2001}).

\bibitem[{\citenamefont{Walsworth et~al.}(2002)\citenamefont{Walsworth, Yelin,
  and Lukin}}]{stop}
\bibinfo{author}{\bibfnamefont{R.~L.} \bibnamefont{Walsworth}},
  \bibinfo{author}{\bibfnamefont{S.~F.} \bibnamefont{Yelin}}, \bibnamefont{and}
  \bibinfo{author}{\bibfnamefont{M.~D.} \bibnamefont{Lukin}},
  \bibinfo{journal}{Opt. Phot. News.} \textbf{\bibinfo{volume}{May Issue}},
  \bibinfo{pages}{50} (\bibinfo{year}{2002}), \bibinfo{note}{and refs therein}.

\bibitem[{\citenamefont{Harris and Sokolov}(1998)}]{harris98}
\bibinfo{author}{\bibfnamefont{S.~E.} \bibnamefont{Harris}} \bibnamefont{and}
  \bibinfo{author}{\bibfnamefont{A.~V.} \bibnamefont{Sokolov}},
  \bibinfo{journal}{Phys.~Rev.~Lett.} \textbf{\bibinfo{volume}{81}},
  \bibinfo{pages}{2894} (\bibinfo{year}{1998}).

\bibitem[{\citenamefont{Hakuta et~al.}(1992)\citenamefont{Hakuta, Marmet, and
  Stoicheff}}]{hakuta92}
\bibinfo{author}{\bibfnamefont{K.}~\bibnamefont{Hakuta}},
  \bibinfo{author}{\bibfnamefont{L.}~\bibnamefont{Marmet}}, \bibnamefont{and}
  \bibinfo{author}{\bibfnamefont{B.~P.} \bibnamefont{Stoicheff}},
  \bibinfo{journal}{Phys.~Rev.~A} \textbf{\bibinfo{volume}{45}},
  \bibinfo{pages}{5152} (\bibinfo{year}{1992}).

\bibitem[{\citenamefont{Lukin et~al.}(1999)\citenamefont{Lukin, Yelin,
  Fleischhauer, and Scully}}]{lukin99}
\bibinfo{author}{\bibfnamefont{M.~D.} \bibnamefont{Lukin}},
  \bibinfo{author}{\bibfnamefont{S.~F.} \bibnamefont{Yelin}},
  \bibinfo{author}{\bibfnamefont{M.}~\bibnamefont{Fleischhauer}},
  \bibnamefont{and} \bibinfo{author}{\bibfnamefont{M.~O.}
  \bibnamefont{Scully}}, \bibinfo{journal}{Phys.~Rev.~A}
  \textbf{\bibinfo{volume}{60}}, \bibinfo{pages}{3225} (\bibinfo{year}{1999}).

\bibitem[{\citenamefont{Ye et~al.}(2002)\citenamefont{Ye, Zibrov, Rostovtsev,
  and Scully}}]{ye02}
\bibinfo{author}{\bibfnamefont{C.~Y.} \bibnamefont{Ye}},
  \bibinfo{author}{\bibfnamefont{A.~S.} \bibnamefont{Zibrov}},
  \bibinfo{author}{\bibfnamefont{Y.~V.} \bibnamefont{Rostovtsev}},
  \bibnamefont{and} \bibinfo{author}{\bibfnamefont{M.~O.}
  \bibnamefont{Scully}}, \bibinfo{journal}{Phys.~Rev.~A}
  \textbf{\bibinfo{volume}{65}}, \bibinfo{pages}{043805}
  (\bibinfo{year}{2002}).

\bibitem[{\citenamefont{Nemeyanov}(1963)}]{nesm63}
\bibinfo{author}{\bibfnamefont{A.~N.} \bibnamefont{Nemeyanov}},
  \emph{\bibinfo{title}{Vapor pressure of chemical elements}}
  (\bibinfo{publisher}{Elsevier, Amsterdam}, \bibinfo{year}{1963}).

\bibitem[{\citenamefont{Sautenkov et~al.}(1999)\citenamefont{Sautenkov, Kash,
  Velichansky, and Welch}}]{saut99}
\bibinfo{author}{\bibfnamefont{V.~A.} \bibnamefont{Sautenkov}},
  \bibinfo{author}{\bibfnamefont{M.~M.} \bibnamefont{Kash}},
  \bibinfo{author}{\bibfnamefont{V.~L.} \bibnamefont{Velichansky}},
  \bibnamefont{and} \bibinfo{author}{\bibfnamefont{G.~R.} \bibnamefont{Welch}},
  \bibinfo{journal}{Laser Physics} \textbf{\bibinfo{volume}{9}},
  \bibinfo{pages}{889} (\bibinfo{year}{1999}).

\bibitem[{\citenamefont{Yelin and Hemmer}(2002)}]{yelin02}
\bibinfo{author}{\bibfnamefont{S.~F.} \bibnamefont{Yelin}} \bibnamefont{and}
  \bibinfo{author}{\bibfnamefont{P.~R.} \bibnamefont{Hemmer}},
  \bibinfo{journal}{Phys.~Rev.~A} \textbf{\bibinfo{volume}{66}},
  \bibinfo{pages}{013803} (\bibinfo{year}{2002}).

\bibitem[{\citenamefont{Fry et~al.}(2000)\citenamefont{Fry, Lukin, Walther, and
  Welch}}]{fry00}
\bibinfo{author}{\bibfnamefont{E.~S.} \bibnamefont{Fry}},
  \bibinfo{author}{\bibfnamefont{M.~D.} \bibnamefont{Lukin}},
  \bibinfo{author}{\bibfnamefont{T.}~\bibnamefont{Walther}}, \bibnamefont{and}
  \bibinfo{author}{\bibfnamefont{G.~R.} \bibnamefont{Welch}},
  \bibinfo{journal}{Opt. Comm.} \textbf{\bibinfo{volume}{179}},
  \bibinfo{pages}{499} (\bibinfo{year}{2000}).

\end{thebibliography}

\end{document}